\documentstyle[12pt]{article}
\begin{document}
\vspace*{1cm}
\begin{center}{\large \bf The Kasteleyn model and a cellular automaton
approach to traffic flow}\\
\vspace{10mm}{\bf J.G.\ Brankov\footnote
{\it Permanent address: Institute of Mechanics,
Bulgarian Academy of Sciences, 1113 Sofia, Bulgaria},
\  V.B.\ Priezzhev} \\
{\it Laboratory of Theoretical Physics, Joint
Institute for Nuclear Research, Dubna, 141980, Russia}\\
\vspace{5mm}{\bf A.\ Schadschneider}\\
{\it Institut f\"ur Theoretische Physik,
Universit\"at zu K\"oln, D-50937 K\"oln, Germany}\\
\vspace{5mm}{\bf M.\ Schreckenberg}\\
{\it Theoretische Physik/FB10, Gerhard-Mercator Universi\"at Duisburg,
D-47048 Duisburg, Germany}
\end{center}
\vspace{10mm} {{\bf Abstract.} We propose a bridge between the
  theory of exactly solvable models and the investigation of traffic
  flow. By choosing the activities in an apropriate way the dimer
  configurations of the Kasteleyn model on a hexagonal lattice can be
  interpreted as space-time trajectories of cars. This then allows for
  a calculation of the flow-density relationship (fundamental
  diagram). We further introduce a closely-related cellular automaton
  model. This model can be viewed as a variant of the
  Nagel-Schreckenberg model in which the cars do not have a velocity
  memory. It is also exactly solvable and the
  fundamental diagram is calculated.}

\hyphenation{one-dimen-sional}

\newpage
\section{Introduction}
In the past years the investigation of traffic flow using cellular
automata, has become quite popular (see e.g.\ \cite{SSNI} and
references therein). In contrast to the hydrodynamical description
\cite{LW,L,KK} this approach is microscopic in the sense that
individual cars can be distinguished similar to the
Follow-the-Leader models \cite{C,PH,BHSSS}. The simplicity of this
approach allows for a number of new applications which cannot be
incorporated easily in the more complex theories of traffic dynamics.
Especially, large scale computer simulations now are possible
\cite{Wolf,Stauff} which are effective even in the case of complicated
street networks.

In this paper we propose a different approach which uses methods
from statistical mechanics. It is well known that a number of exactly
solvable models of statistical mechanics have a graphical
interpretation in terms of (closed) graphs. We suggest to interpret
such graphs as trajectories of cars in traffic. Using the known
partition function of an exactly solvable free-fermion model, the
Kasteleyn dimer model on a hexagonal lattice, we then
are able to calculate the properties of the corresponding traffic
model. In this way we not only obtain informations about the
fundamental diagram (density vs.~flow) but also about the correlation
functions.  Finally, we  discuss the relation of this model to
the approach using cellular automata.

\section{Kasteleyn model}

We suggest and study here a simple statistical two-dimensional lattice
model, the configurations of which can be mapped onto the trajectories
of a discrete (stochastic) one-dimensional traffic problem on a ring
("Indianapolis situation"), see Fig.\ 1. Our traffic model has the
following properties: (i) the cars do not collide, the minimum
distance between any two cars being not less than one lattice spacing;
(ii) the velocities of the cars are (in principle) unlimited.  The
principle difference between the cellular automaton models and the
present approach is that the statistical weights are ascribed globally
to each allowed set of trajectories, rather than determined locally by
the transition probabilities from one space configuration to another.

The statistical model is actually an interpretation of the dimer model
on the hexagonal lattice with anisotropic activities, which has been
suggested by Kasteleyn \cite{K} and studied in a series of papers, see
\cite{NB} and references therein. The hexagonal lattice we represent
as a decorated square lattice with two sites in each vertex shown by a
circle in figure 1a.  We consider the horizontal direction of the
square lattice as the space axis and the vertical one downwards as the
time axis of the traffic problem as usual in traffic theory. The above
interpretation allows us to define an one-to-one mapping of each dimer
configuration on the decorated lattice, Fig.\ 1a, onto a set of
trajectories on the square lattice, Fig.\ 1b. Under that mapping the
circles of the decorated lattice are mapped onto the sites of the square
lattice; a horizontal dimer on the decorated lattice corresponds to a
unit space step of some trajectory, and a vertical dimer corresponds to
a unit time step.  The fact that the Kasteleyn model belongs to
the class of free-fermion models ensures continuity and
non-intersection of the trajectories.  The number of trajectories and
the average velocity of the cars are controlled by the activities $x$
and $t$ of the horizontal and vertical edges of the lattice,
respectively; the activity of the slanted edges in the circles, Fig.
1a, is set to unity. The statistical weight of each allowed set of
trajectories is determined by the total number of horizontal and
vertical unit steps in it. Thus, the generating (or partition)
function of the model on a finite square lattice with $L$ columns
and $M$ rows is given by:

\begin{equation} \label{0}
Z_{L,M}(x,t)= \sum_{\{C\}} x^{N_{x}(C)}t^{N_{t}(C)}
\end{equation}
Here the summation runs over the set of all allowed dimer configurations
on the hexagonal lattice. For periodic boundary conditions,
the solution of the Kasteleyn model in the thermodynamic limit is \cite{NB}

\begin{equation} \label{1}
f_\infty=\lim_{M,N  \to \infty} \frac{1}{M L} \ln Z_{L,M}(x,t)=
\frac{1}{8 \pi^{2}} \int \limits_{\ \ \ 0}^{\ \ \ 2 \pi} \hspace{-2mm} \int
d \alpha d \beta \ln |1-t e^{i\alpha}-x e^{i\beta}|
\end{equation}

The phase diagram of the model has four regions in the
plane $0 \leq t < \infty$,
$0 \leq x < \infty$, \cite {Nagle}, denoted here by $A,B,C$ and $D$.
The phase in region $A$,

\begin{equation} \label{2}
(x,t) \in {\it A}= \{t > 1+x\}
\end{equation}
describes stopped, close-packed cars. There is a trivial empty
phase in region $B$,

\begin{equation} \label{3}
(x,t) \in {\it B} =\{x<1\ ,\ t<|1-x|\}
\end{equation}
and the phase in $C$

\begin{equation} \label{4}
(x,t) \in {\it C}= \{x>1\ ,\ t<|1-x|\}
\end{equation}
corresponds to the zero-density limit of cars moving at infinite speed.
Normal trajectories of cars at all intermediate densities exist in the
region $(x,t) \in D$,

\begin{equation} \label{5}
(x,t) \in {\it D}= \{|1-x| < t < |1+x|\}
\end{equation}
which motivates our interest only in that nontrivial case.

 The integration over $\alpha$ in equation (\ref{1}) can be easily performed
by using the Jensen formula \cite{Jensen}. In region $D$ the result is
\begin{equation} \label{6}
f_{\infty}=\Delta \ln(t)+\frac{1}{2\pi} \int \limits_{\pi \Delta}^{\pi}
d \beta \ln (1-2 x \cos(\beta)+x^{2})
\end{equation}
where

\begin{equation} \label{7}
\Delta=\frac{1}{\pi} \arccos\left(\frac{1+x^2-t^2}{2 x}\right)
\end{equation}

  Let us turn now to the interpretation of the observables in our model.
For a finite lattice under periodic boundary conditions, the total numbers
of steps in time and space can be expressed as

\begin{equation} \label{8}
N_{t}(C)=M N_{a}(C)
\end{equation}
and
\begin{equation} \label{9}
N_{x}(C)=L N_{a}(C) w(C)
\end{equation}
where $N_{a}$ is the number of automobiles in configuration $C$, and
$w(C)$ is the number of windings of each trajectory in the periodic
spatial direction. Therefore, the average density of cars can be defined as

\begin{equation} \label{10}
\rho (x,t)=\frac{1}{LM}\left<N_{t}(C)\right>=
t\frac{\partial}{\partial t}f_{\infty}(x,t)= \Delta(x,t)
\end{equation}

The rigorous definition of the average velocity $v$ of cars is given
by the average value of the ratio of space steps to time steps.
However, as a first approximation we can assume

\begin{equation} \label{11}
v=\frac{\left< N_{x}(C)\right> }{\left< N_{t}(C) \right> }
\end{equation}
The numerator of the above ratio is obtained by differentiation
of expression (\ref{6}) with respect to the activity of the space steps $x$:

\begin{equation} \label{12}
\frac{1}{LM} \left< N_{x}(C)\right>=
x\frac{\partial}{\partial x}f_{\infty}(x,t)=
\frac{1}{2}(1- \Delta (x,t))+ \frac{sign (x-1)}{2}(1-K(x,t))
\end{equation}
Here we have introduced the function
\begin{equation} \label{13}
K(x,t)=\frac{2}{\pi}\arctan \left(\frac{x+1}{|x-1|}
\sqrt{\frac{t^{2}-(1-x)^{2}}{(1+x)^{2}-t^{2}}}\hspace{1mm} \right)
\end{equation}

To obtain the fundamental traffic flow--density diagram, we start with
the definiton of the flow

\begin{equation} \label{14}
q=v\rho
\end{equation}
which, in combination with (\ref{10})-(\ref{12}), gives
\begin{equation} \label{15}
q(x,t)= \frac { \left< N_{x}(C)\right>}{LM}
\end{equation}
The parameter $t$ plays an auxiliary role in our model and we will express
it in terms of the density $\rho$ and activity $x$ by using the relation
$\Delta(x,t) = \rho$, see (\ref{7}) and (\ref{10}). Thus, for the traffic flow
$q(x,\rho)$ we obtain the following explicit expression
\begin{equation} \label{16}
q(x,\rho)=\frac {1-\rho}{2} + \frac{sign (x-1)}{2 \pi}\left[ \pi -
\arccos (\frac{(x-1)^{2}(1+\cos( \pi\rho))}{x^{2}-2x\cos(\pi\rho) +1}
-1)\right]
\end{equation}

   Note that the variable $x$ controls the average velocity $v_m$ of a
single car. Indeed, by considering configurations with only one trajectory,
one can readily see that $v_m = x/t$.

   The flow-density diagram at different fixed values of $x$ is shown in
figure 2. A remarkable feature of that diagram is the existence of two
qualitatively different traffic-flow regimes. For $x < 1$ the flow reaches
its maximum value
\begin{equation} \label{17}
q(x,\rho_{max})=\frac{1}{2}-\rho_{max}
\end{equation}
at the density
\begin{equation} \label{18}
\rho_{max}=\frac{1}{\pi}\arccos(x)
\end{equation}
In the other regime, when $x > 1$, the traffic flow decreases monotonically
with the increase of $\rho$, from $q(x,0) = 1$ to $q(x,1) = 0$.

   The phenomenon of self-organized criticality in the transport flow is
clearly manifested in the asymptotic behaviour of the pair correlation
functions both in space and
time. Obviously, the correlations between time and space steps of the
trajectories in our model correspond to dimer-dimer correlations in the
original model of close-packed dimers on the hexagonal lattice. The latter
can be obtained and studied by using the technique of Fisher and Stephenson
\cite {FS}. In the transport-flow problem one is interested mainly in
the correlations between cars passing a fixed position in space at two
moments of time, say, $t = 0$ and $t = T$, as well as
in the correlations between cars at the same moment
of time, but at a distance $R$ apart. The former case corresponds to
the correlation function for two
vertical dimers in the same row of the decorated square lattice,
and the latter case corresponds to two horizontal dimers in the same column,
see Fig.\ 1. Due to the symmetry of the problem, we  give here the expression
for the temporal correlations only. By using the method of \cite {FS},
for the correlation function $K_{t}(T)$ one obtains in the thermodynamic limit

\begin{equation} \label{19}
K_{t}(T)=-\Biggl|
\frac{1}{(2 \pi)^{2}} \int \limits_{\ \ \ 0}^{\ \ \ 2 \pi} \hspace{-2mm} \int
d \alpha d \beta \frac{e^{i\alpha+iT\beta}}{1-t e^{i\alpha}-x e^{i\beta}}
\Biggr|^{2}
\end{equation}
The  simple  analytical structure of the integrand in the complex
plane  permits the exact evaluation of the above integral
\begin{equation} \label{20}
K_{t}(T)=- \frac{\sin^{2}(\pi T \Delta^{'})}{(\pi x T)^{2}}
\end{equation}
Here $\Delta^{'} \equiv \Delta^{'}(x,t) = \Delta(t,x)$.

    Similarily, for the spatial correlation function we obtain the asymptotic
behaviour $K_{s}(R) \propto R^{-2}$ at large $R$. The power-law
decay of both temporal and spatial correlation functions reflects the
phenomenon of self-organized criticality in our model. However, the following
questions arise: (1) are the obtained critical exponents universal for the
class of statistical models of traffic; (2) will they be the same for the
cellular automaton models.

\section{Comparison with the  cellular automaton \\ approach}

The interpretation of the above results as a cellular automaton with
local rules is not straightforward. The summation over all allowed
dimer configurations corresponds to a global analysis of the
statistics and there are no simple probabilistic {\em local} rules
yielding the same distribution as the partition function (1). But it
is possible to define a model with very similar properties as is shown in
the following.

Consider a probabilistic process according to the following rules.
One time step consists an update of the positions of all cars in a
parallel way. If a car has $n$ empty sites in front of it (next car at
$n+1$) there are at most $n$ partial movements of one site possible
for this car.  The decision how far the car finally moves depends on
a series of random numbers yielding with probability $p$ a '1' and
with $q=1-p$ a '0'.  The car moves as long as 1's are chosen. When a 0
occurs the movement stops until in the next time step this process
starts anew.  So if the first chosen random number is a 0 the car does
not move at all.  If the series has $l$ succeeding 1's the car ends up
at the $l$-th site (possibly directly behind the car ahead if it does
not move itself in the same time step).

A important feature of these rules is that no memory for the velocity
is needed. On the other hand this can lead to unphysical features
through enormous velocity fluctuations from one time step to the next.
Nevertheless one can define a mean maximum velocity of a free driving
car according to the rules just mentioned which is simply given by
$\bar v =p/q$. This is the average distance a single car moves on a free road
without other cars.

In the following we show that the mean-field ansatz already gives the
exact stationary state of the cellular automaton for any system size $L$ and
number of cars $N$. The exactness of the mean-field solution has in
this case the consequence that all possible states of the system are
equally probable.

We characterize a state of the system at time $t$ only by the number $d_i$
($i=1,\ldots,N$) of free sites between car $i$ and car $i+1$ since no
other quantities are needed. For the ring geometry this means that
$d_N$ is the distance between car $N$ and car $1$. It is then rather
simple to write down the master-equation for the time evolution of the
system

\begin{equation}
  P_{t+1}(\{d_i\}) = \sum^{d_N}_{s_1=0}\sum^{d_1}_{s_2=0}
  \cdots\sum^{d_{N-1}}_{s_N=0}\prod^N_{i=0} \left[ p^{s_i}\left(
  \delta_{d_i,s_{i+1}}
  +(1-\delta_{d_i,s_{i+1}})q\right)\right]P_t(\{d_i-s_i+s_{i+1}\}).
\label{master}
\end{equation}

The variable $s_i$ denotes the distance car $i$ drives in the timestep
$t\to t+1$.  The Kronecker-$\delta$ takes into account the fact
that if car $i$ ends up directly behind the next car $i+1$ (i.e.\ $s_{i+1}
= d_{i}$) it does not have the 'deceleration' factor $q$. This is the
main difference to the dimer-model where one has in any case this
factor $q$ (i.e.\ $t$ in the other notation).

The static solution of (\ref{master}) is simply given by $P(\{d_i\})
\equiv 1/{L\choose N}$, i.e.\ asymptotically all probabilities are
equal. This can be seen easily from
\begin{equation}\label{ident}
  1 = \sum^{d_N}_{s_1=0}\sum^{d_1}_{s_2=0}
  \cdots\sum^{d_{N-1}}_{s_N=0}\prod^N_{i=0} \left[ p^{s_i} \left(
  \delta_{d_i,s_{i+1}} +(1-\delta_{d_i,s_{i+1}})q\right)\right]
\end{equation}
which follows from the identity
\begin{equation}
1 = \sum^{d_{i-1}}_{s_i} p^{s_i} \left(
  \delta_{d_{i-1},s_i} +(1-\delta_{d_{i-1},s_i})q\right)
\end{equation}
after interchanging product and sums.
The probability $f_n$ for a car to drive exactly $n$ sites in the
thermodynamic limit $N\to\infty$ is given by
\begin{equation}
f_n = (1-\rho)^np^n[(1-\rho)q+\rho]
\end{equation}
since a car can stop by chance or since another car blocks it.
For the flow in the CA one then obtains
\begin{equation}\label{CAflow}
q_{CA}(p,\rho)= \rho \sum_{n=1}^{\infty}nf_n=
\frac{\rho(1-\rho)p}{1-(1-\rho)p}
=\frac{\rho(1-\rho)\bar v}{1+\rho\bar v}
\end{equation}

To compare the flow (\ref{CAflow}) of the cellular automaton
with the flow (\ref{17}), we have to rescale (\ref{17}) with a factor 2
($q(x,\rho)\to 2q(x,\rho)$).
To illustrate this, we consider the limiting case of low car density
and $ x\rightarrow 1 $. The dotted line in Fig.~2 tends to 0.5 when
$\rho \rightarrow 0$. This reflects the fact that each horizontal line
of the lattice can be occupied by the trajectory of the single car or be
empty with equal probability. In the cellular automata approach,the
motion with infinite velocity is forbidden and the case $x \rightarrow
1$ corresponds to the maximal occupation of all horizontal bonds. This
leads to the shift of the dotted line in Fig.~2 to the diagonal of the square.

After the rescaling the maximum of both flows lie on the curve $1-2\rho$.
Therefore we can find the relation between the parameters $p$ and $x$
in the two models by identifying curves which have the same maximum.
This yields the identification
\begin{equation}
p=\frac{1-\frac{2}{\pi}\arccos x}{(1-\frac{1}{\pi}\arccos x)^2}.
\end{equation}
Fig.\ 4 shows $q_{CA}(p,\rho)$ and $q(x,\rho)$ for $p=3/4$ and $x=1/2$.
Although the slopes for small and high densities are different the
overall agreement of the two curves is very good.

In conclusion we considered the Kasteleyn model as a model of traffic
flow.  This model allows to establish a bridge between the theory of
exactly solvable models and the investigation of traffic flow by
cellular automata. By choosing the activities in an appropriate way it
is possible to interpret the dimer configurations as space-time
trajectories of vehicles. The Kasteleyn model can be related to a
modified Nagel-Schreckenberg model in which the cars do not have a
velocity memory. The fundamental diagram of this modified CA can be
obtained exactly as well and shows good agreement with the fundamental
diagram obtained for the Kasteleyn model.

\vfill
\eject

\newpage\noindent
\centerline{{\Large{{\bf Figure Captions}}}}\\[2cm]
\noindent
Figure 1: Representation of the hexagonal lattice as a decorated
quadratic lattice, fig.~1a, and its mapping
onto the square lattice, fig.~1b. The encircled pairs of sites of the
decorated lattice map on a single site of the square lattice, the vertical
(horizontal) dimers on the former lattice map onto space (time)
steps of the trajectories on the latter.\\[0.8cm]
\noindent
Figure 2: The flow $q(x,\rho)$ for different values of $x$. Curves
with $x<1$ lie below the dotted line ($x=1$), curves with $x>1$ above.
The broken line indicates the location of the maximum flow.\\[0.8cm]
\noindent
Figure 3: The flow $q_{CA}(p,\rho)$ for different values of $p$.\\[0.8cm]
\noindent
Figure 4: Comparison of $q_{CA}(p,\rho)$ (dotted line) and $2q(x,\rho)$ for
$p=3/4$ and $x=1/2$.
\vfill\eject
\newpage
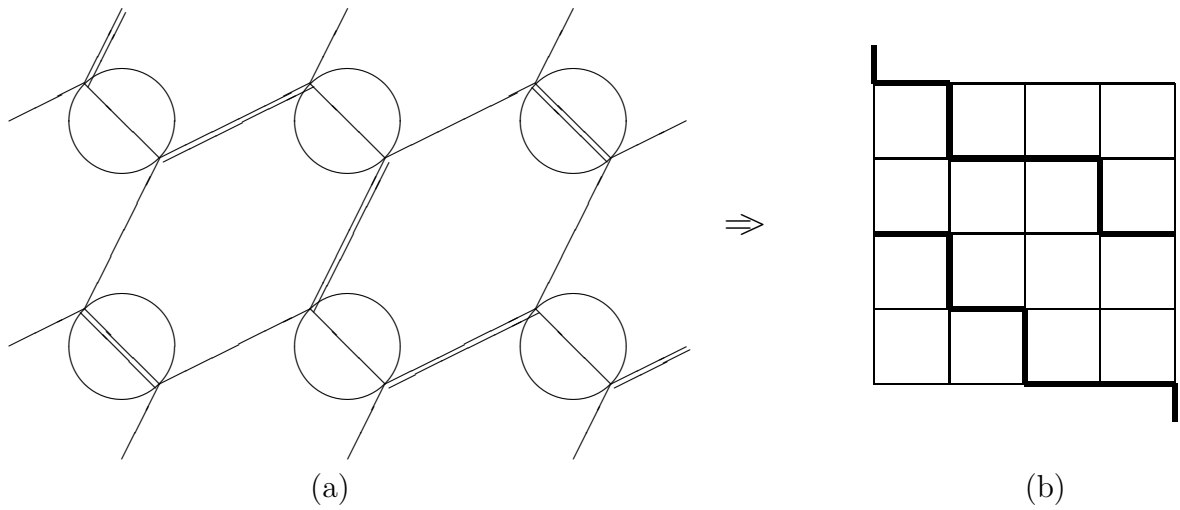
\begin{figure}
\setlength{\unitlength}{1mm}
\begin{picture}(160,70)
\put(15,20){\circle{14.14}}
\put(45,20){\circle{14.14}}
\put(75,20){\circle{14.14}}
\put(15,50){\circle{14.14}}
\put(45,50){\circle{14.14}}
\put(75,50){\circle{14.14}}
\put(115,15){\line(1,0){40}}
\put(115,25){\line(1,0){40}}
\put(115,35){\line(1,0){40}}
\put(115,45){\line(1,0){40}}
\put(115,55){\line(1,0){40}}
\put(115,15){\line(0,1){40}}
\put(125,15){\line(0,1){40}}
\put(135,15){\line(0,1){40}}
\put(145,15){\line(0,1){40}}
\put(155,15){\line(0,1){40}}
\linethickness{2pt}
\put(115,55){\line(1,0){10}}
\put(125,45){\line(1,0){20}}
\put(145,35){\line(1,0){10}}
\put(115,35){\line(1,0){10}}
\put(125,25){\line(1,0){10}}
\put(135,15){\line(1,0){20}}

\put(115,55){\line(0,1){5}}
\put(125,45){\line(0,1){10}}
\put(125,25){\line(0,1){10}}
\put(135,15){\line(0,1){10}}
\put(145,35){\line(0,1){10}}
\put(155,10){\line(0,1){5}}
\thinlines
\put(20,15){\line(2,1){20}}
\put(50,15){\line(2,1){20}}
\put(50.5,14.5){\line(2,1){20}}
\put(80,15){\line(2,1){10}}
\put(80.5,14.5){\line(2,1){10}}
\put(0,20){\line(2,1){10}}
\put(0,50){\line(2,1){10}}
\put(20,45){\line(2,1){20}}
\put(20.5,44.5){\line(2,1){20}}

\put(50,45){\line(2,1){20}}
\put(80,45){\line(2,1){10}}

\put(10,25){\line(1,2){10}}
\put(10,55){\line(1,2){5}}
\put(10.5,54.5){\line(1,2){5}}

\put(15,5){\line(1,2){5}}
\put(45,5){\line(1,2){5}}
\put(40,25){\line(1,2){10}}
\put(40.5,24.5){\line(1,2){10}}

\put(40,55){\line(1,2){5}}
\put(75,5){\line(1,2){5}}
\put(70,25){\line(1,2){10}}
\put(70,55){\line(1,2){5}}

\put(20,15){\line(-1,1){10}}
\put(19.5,14.5){\line(-1,1){10}}
\put(20,45){\line(-1,1){10}}
\put(50,15){\line(-1,1){10}}
\put(50,45){\line(-1,1){10}}
\put(80,15){\line(-1,1){10}}
\put(80,45){\line(-1,1){10}}
\put(79.5,44.5){\line(-1,1){10}}

\put(95,35){\makebox(0,0)[bl]{{\large $ = \hspace{-2mm}>$}}}
\put(40,0){\makebox(0,0)[bl]{(a)}}
\put(135,0){\makebox(0,0)[bl]{(b)}}
\end{picture}
\caption{Representation of the hexagonal lattice as a decorated
quadratic lattice, fig.~1a, and its mapping
onto the square lattice, fig.~1b. The encircled pairs of sites of the
decorated lattice map on a single site of the square lattice, the vertical
(horizontal) dimers on the former lattice map onto  space   (time)
steps of the trajectories on the latter.}
\end{figure}
\
\vfill\eject
\includegraphics{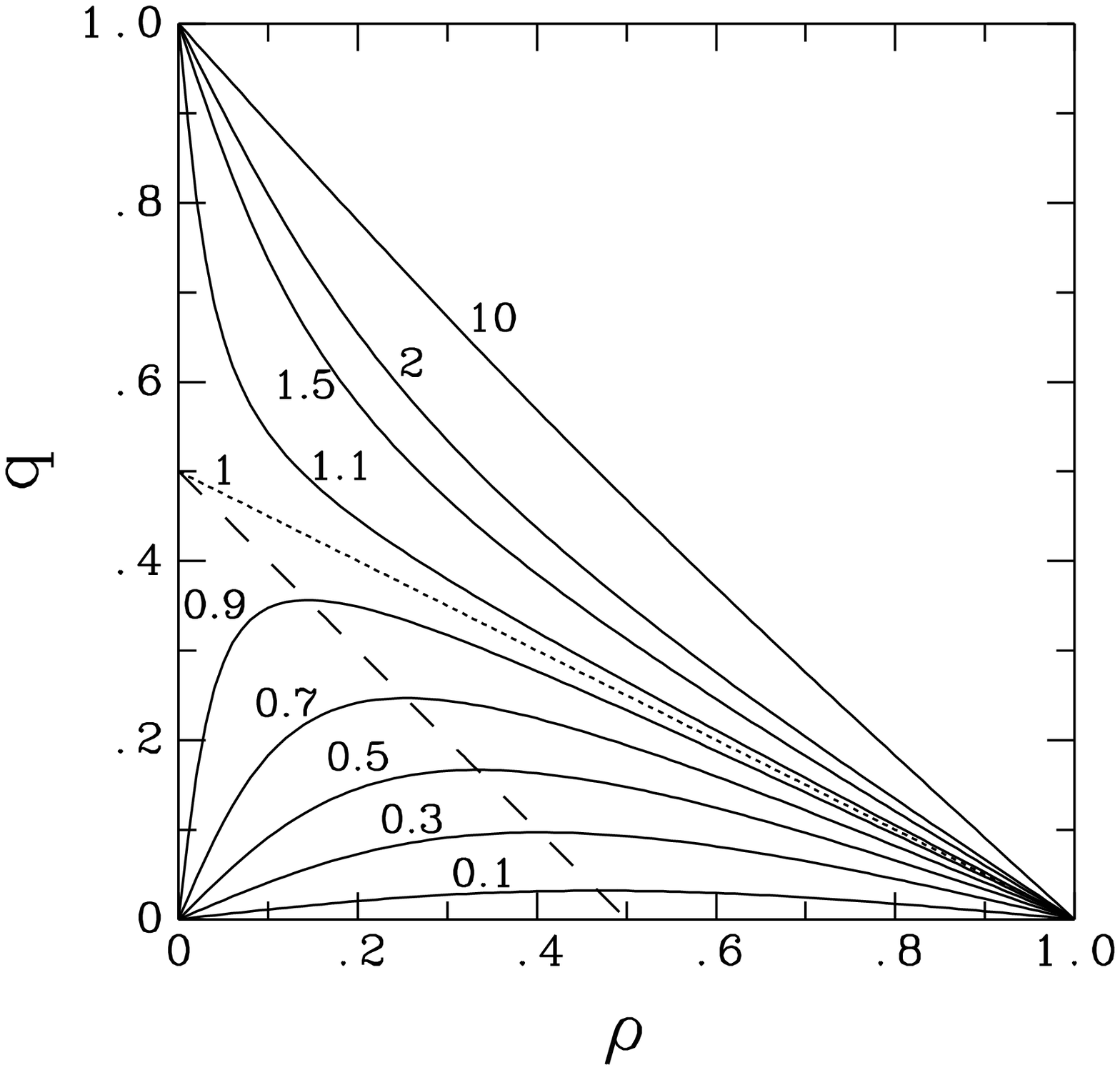}
\noindent
Figure 2: The flow $q(x,\rho)$ for different values of $x$. Curves
with $x<1$ lie below the dotted line ($x=1$), curves with $x>1$ above.
The broken line indicates the location of the maximum flow.
\vfill\penalty -5000\vglue 8.5cm \vfill\eject
\includegraphics{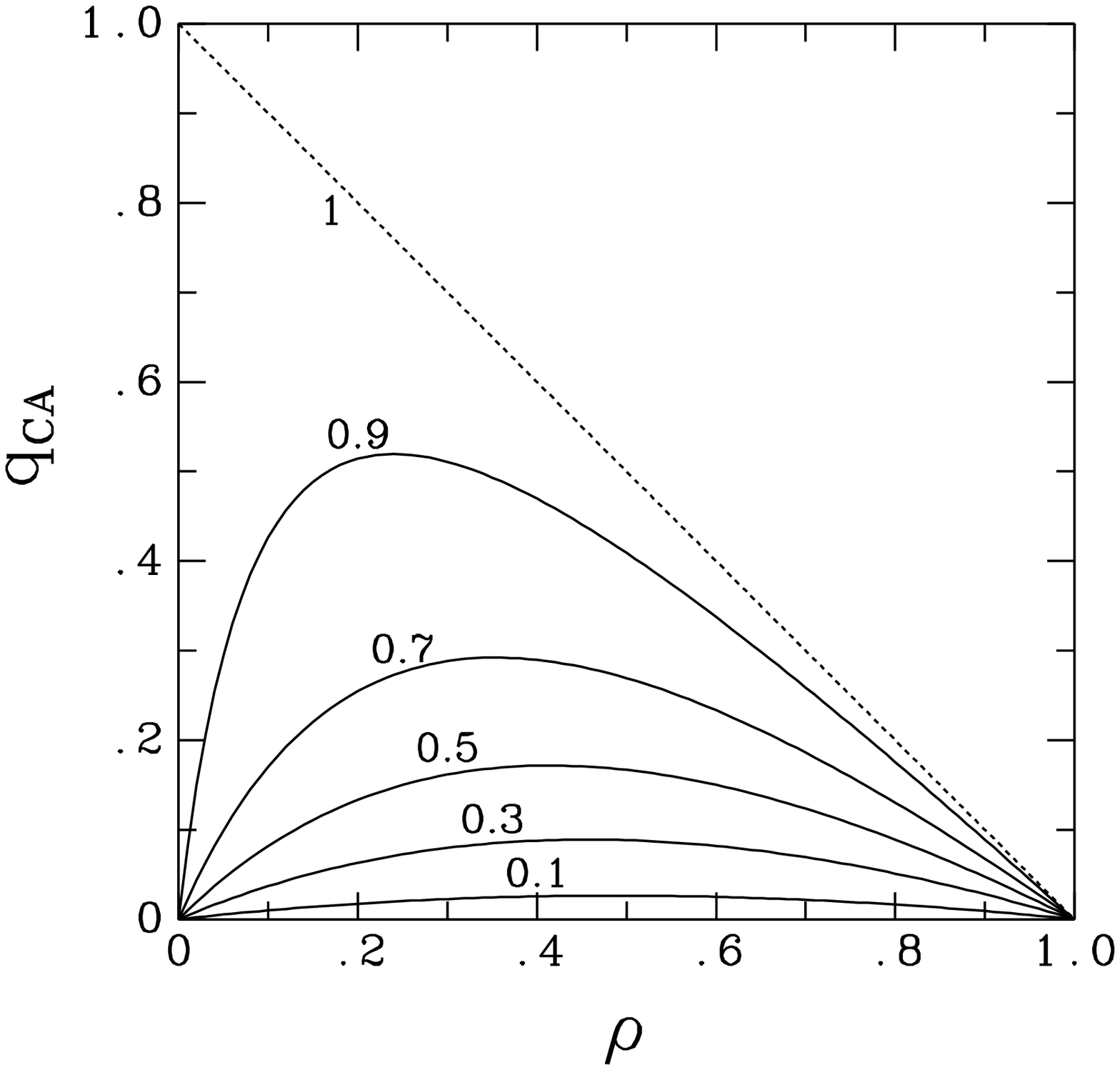}
\noindent
Figure 3: The flow $q_{CA}(p,\rho)$ for different values of $p$.
\vfill\penalty -5000\vglue 8.5cm \vfill\eject
\includegraphics{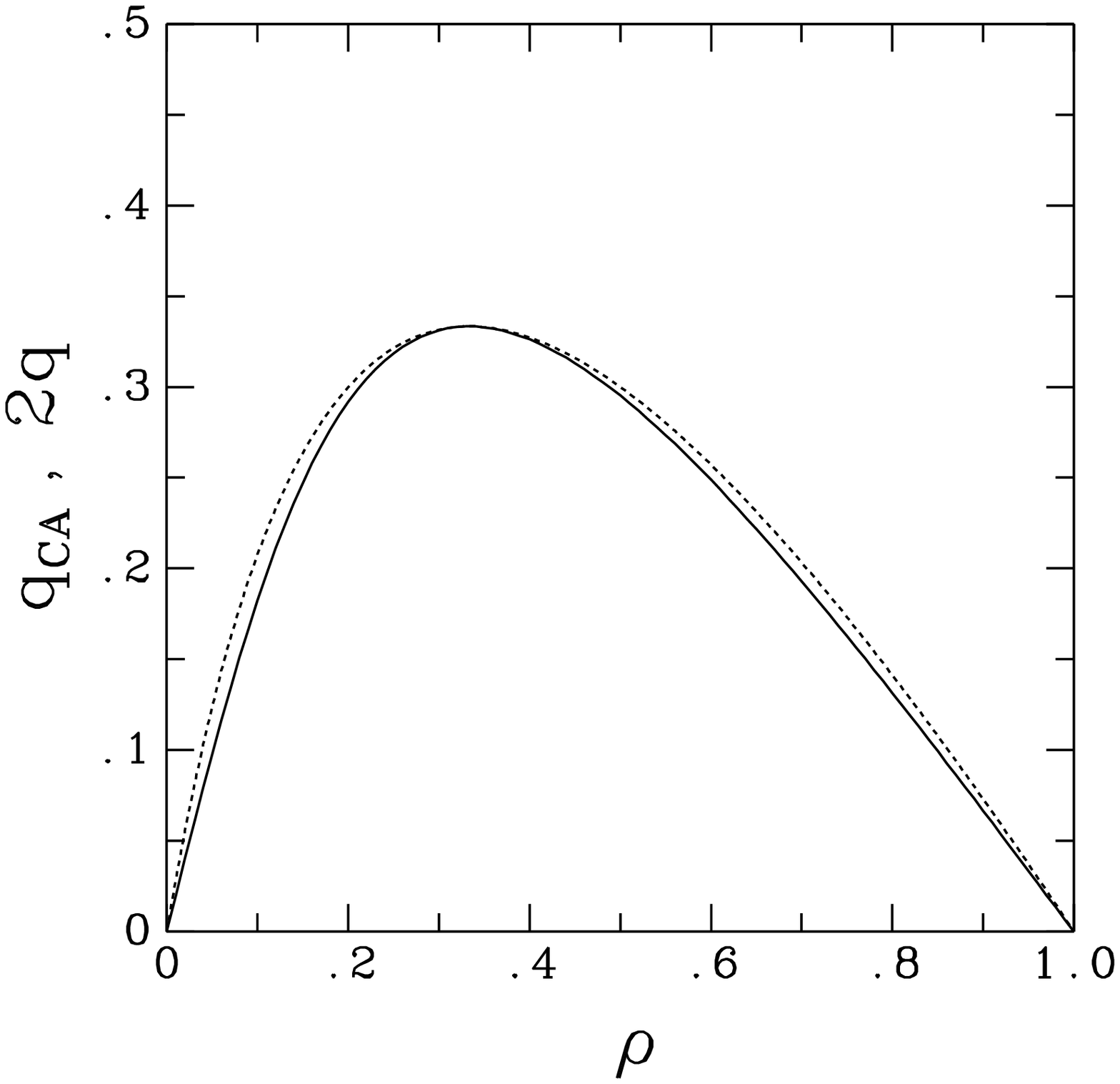}
\noindent
Figure 4: Comparison of $q_{CA}(p,\rho)$ (dotted line) and $2q(x,\rho)$ for
$p=3/4$ and $x=1/2$.
\vfill\penalty -5000\vglue 8.5cm
\vfill\eject

\end{document}